      [1999/12/01 v1.4c Il Nuovo Cimento]
\documentclass{cimento}
\usepackage{graphicx}
\title{Model Independent Methods of Describing GRB Spectra Using BATSE MER Data}
\author{P.~Veres\from{ins:x}\from{ins:x1}\ETC,
I.~Horv\'ath\from{ins:y},
Z.~Bagoly\from{ins:x},
L.G.~Bal\'azs\from{ins:a},
A.~M\'esz\'aros\from{ins:b}\from{ins:d},
G.~Tusn\'ady\from{ins:c}
	\atque 
F.~Ryde\from{ins:d}\from{ins:e}}
\instlist{\inst{ins:x} E\"{o}tv\"{o}s University, H-1117 Budapest, P\'azm\'any P. s.  1./A, Hungary.
\inst{ins:x1} Department of Physics, University College Cork, Ireland.
\inst{ins:y} Dept. of Physics, Bolyai Military University, Budapest, Box-12, H-1456, Hungary.
\inst{ins:a}Konkoly Observatory, H-1525 Budapest, POB 67, Hungary.
\inst{ins:b}Astronomical Institute of the Charles University, Prague, Czech Republic.
\inst{ins:c}R\'enyi Institute of Mathematics, H-1364 Budapest, POB 127, Hungary.
\inst{ins:d}Department of Physics, The Royal Institute of Technology, 10691 Stockholm, Sweden.
\inst{ins:e}Stockholm Observatory, AlbaNova, SE-106 91 Stockholm, Sweden.}
\PACSes{\PACSit{00.00}{By the way, which PACS is it, the 00.00? GOK.}
\PACSit{---.---}{\ldots}}
\begin{document}

\maketitle

\begin{abstract}
The Gamma Ray Inverse Problem is discussed. Four methods of spectral deconvolution are presented here and applied to the BATSE's MER data type. We compare these to the Band spectra.
\end{abstract}
\section{Introduction}

The Gamma Ray Inverse Problem (\cite{ref:loredo}) is as follows: from a given number of observed parameters we have to reconstruct the intrinsic spectra. This will be estimated at some other number of points. The relation between the measured parameters and the intrinsic spectra is given by the detector response matrix (DRM). In the case of the BATSE MER data, the spectra has to be approximated at $62$ intervals from $16$ measured count rates. \\

The inverse problem corresponds to the discretized version of the Fredholm integral equation of the first kind. 
\begin{eqnarray}
d(y)=\int  R(x,y) f(x) dx\label{alt-2}.
\end{eqnarray}
Here $R(x,y)$ is the so-called kernel function, in our case the detector's response matrix. $d(y)$ is the measured count rate, $f(x)$ is the quantity to be determined. This is known to be an ill-posed and underdetermined problem. 
In the case of BATSE MER $d$ is a $16$ element vector, $f$ a $62$ element vector and $R$ is a $16\times 62$ matrix.

\section{Data}
Data were taken from the BATSE database (ftp://cossc.gsfc.nasa.gov). We only used the Medium Energy Resolution data type. Background fits were made using intervals by Norris (http:// cossc.gsfc.nasa.gov/ docs/ cgro/ \-batse/ batseburst/ sixtyfour\_ms/ bat\_files.revamp\_join) for the \textit{cat64} data. We have subtracted $3^{rd}$ degree polynomials from all the $16$ channels of MER data.
\section{Methods}
Throughout the literature there are articles which discuss in detail the inverse problem in astronomy (\cite{ref:loredo,ref:bouchet}) and other sciences (\cite{ref:parker}). 
Gamma-ray bursts' spectra are most commonly reconstructed by the method parameter fitting also knows as forward folding \cite{ref:preece}. This is done by postulating spectral forms with some adjustable parameters. Once multiplied with the DRM we get a solution for the count rates and we then compare them to the measured set of data. We accept the set of parameters which minimizes the $\chi^2$ of the solution and the measured counts. Although widely used and a well established method, still it is obliging when it comes to the shape of the spectra and it consists of estimating the parameters and not the spectra itself \cite{ref:loredo}.

\subsection{Singular Value Decomposition}

We decompose the response matrix using this well known mathematical method. We solve the GRIP by multiplying with the generalised inverse obtained through SVD. Spectra obtained from simply multiplying with the general inverse tend to be noisy. To avoid this we can use filters. 
\subsection{Maximum entropy method}

This method introduces some \textit{a priori} information to reconstruct the spectra ($m$). This is done by fitting a Band GRB spectral form \cite{ref:band}. We propose to maximize the expression: $L(f,\lambda)=\lambda S(f,m)-\frac 1 2 \chi^2$
where $\lambda$ is a parameter, $m$ is the \textit{a priori} spectra (from model fitting or SVD). It is with respect to this that we measure the entropy of the intrinsic spectra ($f$). The expression of the entropy is as follows: $S(f,m)=\sum_{j=1}^{N(=62)}f_j-m_j-f_j \log \left(\frac{f_j}{m_j}\right)$. When $L$ has a maximum then $\nabla L(f,\lambda)=0.$. From this we derive: $ \lambda \log \left(\frac{f}{m}\right)= R^\top [\sigma ^{-2}] (d-R f)$, where $[\sigma ^2]$ is a diagonal matrix with elements of the $\sigma ^2$ vector and $R$ is the response matrix. Since $\log \left(\frac{f}{m}\right)$ results from the linear combination of the $R^\top$'s columns, we can search for the solution in the form of: $f=m e^ {R^\top w}$. Here $w$ is a $16$ element vector. This ensures that $f$ will always be positive, and since $w$ has a smaller scale, it is easier to find it as a result of a minimizing process. The choice of $\lambda$ can be made with the so-called cross-validation technique or another method based on a Bayesian approach. We have applied the method to BATSE MER data. First results suggest that the Band GRB function provides a good enough fit.
\subsection{Backus Gilbert method}
\begin{figure}
\begin{center}
\includegraphics[width=0.5\linewidth]{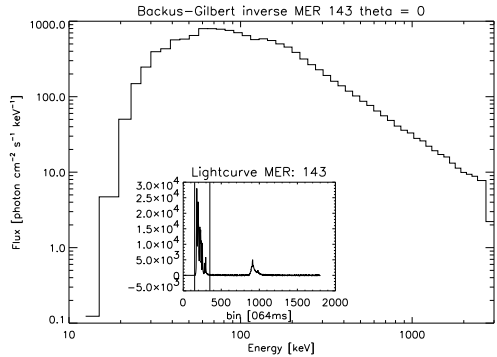}
\caption{Sample Backus Gilbert inverse.} \label{fig:1}
\vspace{-1cm}
\end{center}
\end{figure}

Its main advantage is that it is a linear method(\cite {ref:backus}), data in the intrinsic spectra is reconstructed as a linear combination of the measured count rates: $f(E_j)=\sum_{i=1}^M a_i(E_j)d_i=\sum_{k=1}^N \Delta_k(E_j) f_k$ and $\Delta_k(E_j)=\sum_{i=1}^M a_i(E_j) R_{i,k}.$. We call $\mathcal R (E)$ the resolution function. Our aim is to approximate the resolution function with Dirac-delta function as well as possible. By doing this we minimize for $a_i$. The other issue at hand is the variance of the spectra. This is dealt with by minimizing $\sigma_j^2(E_j)=\sum_{i=1}^M a^2_i(E_j)\sigma^2_i$ also for $a_i$. These two expressions are combined into a single one depending on the parameter $\theta$ in the following manner: $w(E_j,\theta)=\mathcal R (E_j) \cos \theta +v \sigma_j^2(E_j) \sin \theta$. By minimizing $w$ with respect to $a_i$ we get the Backus Gilbert inverse (Fig.\ref{fig:1}). ($v$ is merely a factor to get the two components to the same order of magnitude)
\subsection{Phillips-Towmey method}

We will provide only a brief description of the method. Like at the maximum entropy method, we are also in need here of an \textit{a priori} spectrum. This information is used to reduce the characteristic length scale of the problem. We introduce the sum-of-squares of the spectrum's $k$-th derivative ($C_k$).
%

We propose to minimize some linear combination of the $\chi ^2$ and the $C_k$. As an \textit{a priori} spectrum we fit a Band GRB function. We find that there is only insignificant gain in the function to be minimized with respect to the Band GRB function.

\section{Conclusion}
Having applied several of the above presented methods to invert BATSE MER spectra, we conclude that the methods give a consistent estimation of the incident spectrum each with regard to the minimizing function they use. The SVD method is the fastest, the Backus Gilbert requires the most time, but it is less sensitive to noise. We have also found that the Band GRB function provides a good fit to the spectra we analyzed.
\acknowledgments This research was supported by  OTKA grant
T048870 and by a grant from Swedish Wenner-Gren Foundations (A.M.).

\end{document}